\documentclass[a4paper,12pt]{article}
\usepackage{amsmath,amsfonts,epsfig}
\numberwithin{equation}{section}
\usepackage{slashed}

\def\bra{\langle}
\def\ket{\rangle}

\def\tr{\mathrm{tr}}

\def\beq{\begin{equation}}
\def\eeq{\end{equation}}

\def\2b2[#1,#2][#3,#4]{\left( \begin{array}{cc} #1 & #2 \\ #3 & #4 \end{array} \right)}
\def\3b3[#1,#2,#3][#4,#5,#6][#7,#8,#9]{\left( \begin{array}{ccc} #1 & #2 #3 \\ #4 & #5 & #6\\#7&#8&#9\end{array} \right)}

\newcommand{\C}[1]{\mathcal{#1}}

\def\li{\mathrm{Li}}

\def\Mm{M_{\mathrm{mess}}}

\def\ov{\overline}

\author{K.~Benakli\footnote{kbenakli@lpthe.jussieu.fr} and M.~D.~Goodsell\footnote{goodsell@lpthe.jussieu.fr}}

\date{}
\title{Dirac Gauginos in General Gauge Mediation}

\begin{document}

\maketitle
\vspace{-1cm}
\begin{center}
\emph{Laboratoire de Physique Th\'eorique et Hautes Energies\footnote{Unit\'e Mixte de Recherche du CNRS UMR 7589}, CNRS, UPMC Univ Paris 06, Boite 126,
4 place Jussieu, F-75252 Paris Cedex 05 France}
\end{center}
\abstract{We extend the formulation by Meade, Seiberg and Shih of general gauge mediation of supersymmetry breaking to include Dirac masses for the gauginos. These appear through mixing of the visible sector gauginos with additional states in  adjoint representations. We illustrate the method by reproducing the existing results in the literature for the gaugino and sfermion masses when preserving R-symmetry. We then explain how the generation of same sign masses for the two propagating degrees of freedom in the adjoint scalars can be achieved. We end by commenting on the use of the formalism for describing $U(1)$ mixing.}

\section{Introduction}

The study of patterns of supersymmetry breaking is of central importance as it is a necessary step in trying to understand the possible connections between non-supersymmetric infrared vacua, that might be useful in describing the world at accessible energies, to supersymmetric theories, that might describe a fundamental theory in  the ultraviolet. Among possible descriptions of the breaking, gauge mediation scenarios \cite{Dine:1981za,Dimopoulos:1981au,Nappi:1982hm,AlvarezGaume:1981wy,Dimopoulos:1982gm,Affleck:1984xz,Dine:1993yw,Dine:1994vc,Dine:1995ag,Dine:1996xk,Giudice:1998bp} 
are of special interest because of their simplicity,  and because of certain phenomenological features, such as the universality of scalar soft masses which  provides a solution to the SUSY flavor problem. Recently, Meade, Seiberg and Shih (MSS) \cite{Meade:2008wd} proposed a framework called ``General Gauge Mediation'' (subsequently considered in \cite{Ibe:2008si,Endo:2008gi,Carpenter:2008wi,Ooguri:2008ez,Distler:2008bt,Intriligator:2008fr,Dudas:2008eq,Craig:2008vs,Jager:2008fc,Carpenter:2008rj,Intriligator:2008fe,Seiberg:2008qj,Csaki:2008sr}) 
aimed at going beyond the ordinary models with perturbative secluded sectors to incorporate  the case of arbitrary SUSY breaking sectors. Following MSS gauge mediation is defined as the ability to decouple the theory into a visible sector and a separate hidden sector (where SUSY is broken) when taking the limit of all the visible gauge couplings to zero.

While a definitive requirement for viable supersymmetry breaking is the generation of masses for the gauginos, the nature of these masses is not fixed. It can be either of Majorana type, which is the case in the MSSM, or of Dirac type which is necessary if $R$-symmetry remains unbroken. Investigations for a microscopic origin of the Dirac gaugino masses have  been motivated by the possibility of 
 supersoft  mass terms \cite{Polchinski:1982an,Fox:2002bu, Chacko:2004mi}, 
brane models with non-supersymmetric intersections (non-supersymmetric fluxes) \cite{Antoniadis:2005em, Antoniadis:2006eb, Antoniadis:2006uj}, or the possibility of using calculable $R$-symmetric $F$-term SUSY breaking  models\cite{Hall:1990hq, Kribs:2007ac, Amigo:2008rc}. From an effective low energy point of view,  the expected minimal extension of the MSSM to include Dirac masses is explicitly summarized in \cite{:2008gva} and corresponding experimental signature of Dirac gauginos studied in  \cite{Choi:2008pi, Nojiri:2007jm}.

This work deals with the possibility of describing the above class of models with Dirac gauginos using the framework of MSS, i.e. we determine the general consequences of coupling a chiral multiplet in the adjoint representation of each of the MSSM gauge groups to the messenger sector. Although these kind of models  may appear to be a form of direct mediation, we assume, as done in the above previous literature with Dirac gaugino masses, that there is no superpotential term coupling the adjoint fields to the visible matter superfields. Therefore, the mixing between the adjoint fermions and gauginos vanishes at zero gauge coupling, and we can extend the MSS formalism. This is a natural assumption, since the only dimension four operators respecting the gauge symmetries that could couple the adjoint fields to the MSSM involve only the Higgs. Provided that we use the extended MSS definition \cite{Meade:2008wd} of General Gauge Mediation, that we define such couplings to vanish when we also take the gauge couplings to zero, these fall into the MSS formalism. Such operators can be  desirable for phenomenological reasons, for instance to provide a decay channel to the bino. However, the Higgs sector presents challenges  for gauge mediation, especially because of the difficulty in generating both $\mu$ and $B\mu$ terms with the right magnitude. We will not discuss these issues further, nor others such as gauge unification, in this work and refer to the previous literature cited above, and also \cite{Bachas:1995yt,Han:1998pa,Ibe:2007ab,Porto:2007ed}.

Section \ref{REVIEW} reviews the formulation of MSS, that uses current correlators in order to parametrize the visible sector soft terms. Section \ref{FORMALISM} explains the necessary extensions in order to include the Dirac gaugino masses and gives the corresponding formulae for the sfermion masses. Section \ref{APPLICATIONS} illustrates the use of current correlations to reproduce the result of existing models and extend them to the case of both a D and an F-term. Previous microscopic derivations of  supersoft Dirac gaugino masses were plagued with a problem: in the perturbation expansion one obtains a tachyonic mass for one component of the scalars in the adjoint representation \cite{Fox:2002bu, Chacko:2004mi, Antoniadis:2006uj}. Section \ref{SCALARS} explains how this is fixed in the case of multiple messengers by an appropriate choice of superpotential. Some examples are given to illustrate the results. Finally, section \ref{MIXING} provides a short discussion on the use of current correlators for describing kinetic mixing between two Abelian gauge vectors.

\section{Quick Review of General Gauge Mediation}
\label{REVIEW}

The philosophy of general gauge mediation \cite{Meade:2008wd} (see also \cite{Ooguri:2008ez, Distler:2008bt, Intriligator:2008fr}) is to treat the messenger sector as coupling to a supersymmetric visible sector (presumably the MSSM, but not necessarily) linearly via currents to the gauge superfields. Correlators of these currents then determine the observable data, which consist of the scalar and gaugino masses. These are then specified by three real and three complex parameters respectively, and satisfy two sum rules. 

The gauge superfield $\C{V}$ couples to the gauge current superfield of the messenger sector $\C{J}$ where
\beq
\C{J} = J + i \theta j - i \ov{\theta}\bar{j} - \theta \sigma \bar{\theta} j_\mu + \frac{1}{2} \theta \theta \bar{\theta} \bar{\sigma}^\mu \partial_\mu j - \frac{1}{2} \bar{\theta}\bar{\theta} \theta \sigma^\mu \partial_\mu \bar{j} - \frac{1}{4}\theta\theta \bar{\theta} \bar{\theta} \square J
\eeq
via a linear term
\beq
\C{L}_{int} = 2g \int d^4 \theta \C{J} \C{V} = g(JD - \lambda j - \bar{\lambda} \bar{j} - j^\mu V_\mu).
\eeq
By defining correlators 
\begin{align}
\bra J (p) J (-p) \ket &= \tilde{C}_{0} (p^2/M^2; M/\Lambda) \nonumber \\
\bra j_{\alpha} (p) \bar{j}_{\dot{\beta}} (-p) \ket &=  -\sigma^{\mu}_{\alpha \dot{\beta}} p_\mu \tilde{C}_{1/2} (p^2/M^2; M/\Lambda) \nonumber \\
\bra j_{\mu} (p) j_{\nu} (-p) \ket &= -(p^2 \eta_{\mu \nu} - p_\mu p_\nu) \tilde{C}_{1} (p^2/M^2; M/\Lambda) \nonumber \\
\bra j_{\alpha} (p) j_{\beta} (-p) \ket &= \epsilon_{\alpha \beta} M \tilde{B}_{1/2} (p^2/M^2)
\label{MSSCorrelators}
\end{align}
where $M$ is a mass scale defined by the correlators and $\Lambda$ is the cutoff for the theory, we find
\beq
\tilde{C}_a (p^2/M^2; M/\Lambda) = 2\pi^2 c \log \Lambda/M + \tilde{C}_a (p^2/M^2)
\eeq
where $c$ is independent of $a$ and determines the change in beta function due to the messenger matter; and we can calculate terms in the effective Lagrangian
\beq\begin{split}
\delta \mathcal{L}_{eff} =& \frac{1}{2} g^2_r \tilde{C}_{0} (0) D^2  - g^2  i \tilde{C}_{1/2} (0) \lambda \sigma^{\mu}\partial_{\mu} \bar{\lambda} \\
&-  \frac{1}{4} g^2\tilde{C}_{1} (0) F_{\mu \nu}F^{\mu \nu} - \frac{1}{2} g^2 (M \tilde{B}_{1/2} (0) \lambda \lambda + c.c.) .
\end{split}\eeq
We can then read off the Majorana gaugino masses:
\beq
m_{M} = g^2 M \tilde{B}_{1/2} (0).
\eeq
The scalar mass for a sfermion $\tilde{f}$ in representation $f$ of gauge group $r$ with coupling $g_r$ is given by
\beq
m_{\tilde{f}}^2 = g_1^2 Y_f \xi + \sum_{r=1}^3 g_r^4 C_2 (f;r) A_r,
\label{MSSsfermion}
\eeq
where $Y_f$ is the associated hypercharge, $C_2 (f;r)$ is the quadratic Casimir and $\xi$ is a Fayet-Ilioupoulos term, and 
\beq
A_r = - \int \frac{d^4 p}{(2\pi)^4} \frac{1}{p^2} \bigg(3 \tilde{C}^{(r)}_1 (p^2/M^2)  - 4 \tilde{C}^{(r)}_{1/2} (p^2/M^2) + \tilde{C}^{(r)}_0 (p^2/M^2)\bigg) .
\eeq
Thus the gaugino masses are a priori unrelated to the sfermion masses and the change in beta function. However, there are two sum rules obeyed by the scalars, since $U(1)_Y$ and $U(1)_{B-L}$ are non-anomalous:
\beq\begin{split}
\tr (Ym^2) - g_1^2 \xi \tr Y^2 &= 0\\
\tr \bigg((B-L)m^2\bigg) - g_1^2 \xi \tr \bigg((B-L)Y\bigg) &= 0.
\end{split}\eeq

\section{Adjoints and General Gauge Mediation}
\label{FORMALISM}

Our aim is to see how additional matter in the adjoint representation of the standard model gauge groups fits within the context of general gauge mediation. 
For non-Abelian groups adjoint chiral fields $\C{X} (= X + \sqrt{2} \theta \chi + \theta \theta F_X + ...)$  couple to the gauge fields via the current
\beq
\C{J}_X = [ \C{X} , \C{X}^{\dagger}] 
\label{JX}\eeq
and one could attempt to parametrise their behaviour entirely in this way. However, we find the prescription lacking when we want to describe the gaugino masses: the adjoint fermions may mix with the gauginos via Dirac mass terms. To include them, we consider the coupling of the adjoint to messenger fields via the superpotential
\beq
W \supset \lambda_X \C{X} \C{J}_2 
\eeq
leading to a coupling in the Lagrangian
\beq
\C{L} \supset \lambda_X \bigg[ \chi j_2 + \ov{\chi} \ov{j}_2 + X F_2 + \bar{X} \bar{F}_2 \bigg] .
\eeq
Here $\C{J}_2 (= J_2 + \sqrt{2} \theta j_2 + \theta \theta F_2 + ...)$ is a chiral superfield. 
Superficially this may appear to be a form of direct mediation, but as explained in the introduction, since we are not considering the Higgs sector it is natural to assume that there is no superpotential term coupling the adjoint fields to the visible matter. In this way, the mixing between the adjoint fermions and gauginos vanishes at zero gauge coupling, and we can extend the MSS formalism.
The calculation of this mixing involves,  in addition to the correlator of $j_\alpha$ (the chiral fermionic component of the messenger gauge current) with itself, the new correlators 
\begin{align}
\bra j_{2\alpha} (p) \bar{j}_{2\dot{\beta}} (-p) \ket &= -\sigma^{\mu}_{\alpha \dot{\beta}} p_\mu \tilde{E}_{1/2} (p^2/M^2; M/\Lambda) \nonumber \\
\bra j_{\alpha} (p) \bar{j}_{2\dot{\beta}} (-p) \ket &= -\sigma^{\mu}_{\alpha \dot{\beta}} p_\mu \tilde{G}_{1/2} (p^2/M^2;M/\Lambda) \nonumber \\
\bra j_{\alpha} (p) j_{2\beta} (-p) \ket &= \epsilon_{\alpha \beta} M \tilde{H}_{1/2} (p^2/M^2) \nonumber \\
\bra j_{2\alpha} (p) j_{2\beta} (-p) \ket &= \epsilon_{\alpha \beta} M \tilde{I}_{1/2} (p^2/M^2) 
\label{fermion_correlators}\end{align}
(here $M$ is a mass scale that is defined through the above correlators) leading to terms in the effective Lagrangian  
\beq\begin{split}
\delta \mathcal{L}_{eff} =& - g \lambda_X  M\tilde{H}_{1/2} (0) \chi \lambda - g \lambda_X  i \tilde{G}_{1/2} (0) \chi \sigma^{\mu}\partial_{\mu} \bar{\lambda} \\
&- \frac{1}{2} \lambda_X^2 M \tilde{I}_{1/2} (0) \chi \chi - \frac{1}{2}\lambda_X^2 i\tilde{E}_{1/2} (0)\chi \sigma^{\mu}\partial_{\mu} \bar{\chi} + c.c.
\end{split}\eeq
 Then, allowing for a possible tree level majorana mass term for the adjoint fermions of $\frac{1}{2}\hat{M}^0_X M (\chi \chi + \ov{\chi} \ov{\chi})$ (where $\hat{M}^0_X$ is dimensionless), we find to the lowest order in $g$ and $\lambda_X$ two majorana fermions of mass
\beq\begin{split}
m_{\pm} = \frac{1}{2} M \bigg[ &\hat{M}^0_X + \lambda_X^2 \tilde{I}_{1/2} (0) + g^2 \tilde{B}_{1/2} (0)   \\
&\pm \sqrt{ \bigg( \hat{M}^0_X + \lambda_X^2 \tilde{I}_{1/2} (0) - g^2 \tilde{B}_{1/2} (0)\bigg)^2  + 4 g^2 \lambda_X^2 \tilde{H}_{1/2}^2 (0)} \bigg]
\end{split}\eeq   
In the case (such as when preserving R-symmetry) where for the majorana terms $\hat{M}^0_X = 0$ and $\tilde{B}_{1/2} (0) = \tilde{I}_{1/2} (0) = 0$, we recover two majorana fermions of mass $\tilde{H}_{1/2} (0)$ which combine to yield a Dirac gaugino with mass
\beq
m_D = g \lambda_X M \tilde{H}_{1/2} (0).
\label{DiracOnly}\eeq 

Now we turn to the computation of scalar masses. The one-loop generated gaugino soft masses can induce soft masses for the scalars at higher order, as in the case of ordinary gauge mediation. This is manifest in singular behaviour in the correlator $\tilde{C}_{1/2}$ at low momentum; if we compute $\tilde{C}_{1/2}$ at 
two loops we obtain a term of the form
\begin{eqnarray}
 \tilde{C}_{1/2} (p^2) &\supset & - \frac{1}{p^2} \bar{\sigma}^{\mu}_{\dot{\beta} \alpha} p_{\mu}|\lambda_X|^2 \frac{1}{2}\int d^4 x e^{i p \cdot x} \bra j_{\alpha} (x) \bigg(\int d^4 z_1 j_2 \chi \bigg)  \bigg( \int d^4 z_2 \ov{j}_2 \ov{\chi} \bigg) \ov{j}_{\dot{\beta}} (0) \ket \nonumber \\
&\supset& \frac{1}{p^2+(M \hat{M}_X^0)^2} |\lambda_X|^2| \tilde{H} (p) |^2 + ...,
\end{eqnarray}
which is divergent in the infra-red if $\hat{M}_X^0 = 0$. In this case of Dirac gaugino masses, we therefore resum all of the one-particle irreducible contributions to the soft mass-squareds, giving the improved formula 
\beq
A_r = - \int \frac{d^4 p}{(2\pi)^4} \frac{1}{p^2} \bigg(3 \tilde{C}^{(r)}_1 (p^2/M^2)  -\frac{4 \tilde{C}^{(r)}_{1/2} (p^2/M^2)}{1+g_r^2 \tilde{C}^{(r)}_{1/2} (p^2/M^2)}+\tilde{C}^{(r)}_0 (p^2/M^2) \bigg) 
\eeq
which is finite in the infra-red. We then find that the conclusions from general gauge mediation are unchanged: the MSS sum rules are preserved, and the gaugino masses are a priori unrelated to the scalar masses. However, some of the techniques \cite{Distler:2008bt,Intriligator:2008fr} developed to examine the leading order terms in the SUSY breaking order parameter cannot be applied here since the correlators $\tilde{C}_a$ no longer correspond to coefficients in the effective action. 

Note that we did not resum $\tilde{C}_0$ or $\tilde{C}_1$. We must have no pole in $\tilde{C}_{1}$, or we would generate a mass for the gauge bosons, and hence it is not necessary to resum that term. Moreover, if we consider $\tilde{C}_{0}$ with $\hat{M}_X^0 = 0$, we find
\begin{align}
 \tilde{C}_{0} (p^2/M^2) &\supset |\lambda_X|^2 \frac{1}{2} \int d^4 x e^{i p \cdot x} \bra J (x) \bigg(\int d^4 z_1 X F_2 \bigg)  \bigg( \int d^4 z_2 X^\dagger F_2^\dagger \bigg) J (0) \ket \nonumber \\
&\supset |\lambda_X|^2| \bra J(p) F_2 (-p) \ket |^2 \bra X (p) X^\dagger (-p) \ket + ... \nonumber \\
&\supset \frac{1}{p^2} |\lambda_X|^2| \bra J(p) F_2 (-p) \ket |^2 + ...,
\end{align}
but rather than resumming the whole of $\tilde{C}_0$, we find that we should just include further terms in the $\bra X (p) X^\dagger (-p) \ket$ propagator. We also find a contribution from the $\bra X(p) X(-p) \ket$ propagator and its conjugate. These are straightforward to incorporate; defining the correlators
\beq\begin{split}
\bra F_2 (p) F_2^{\dagger} (-p) \ket &=  M^2\tilde{F}_0 (p^2/M^2)  \\
\bra F_2 (p) F_2 (-p) \ket &=  M^2\tilde{F}_0^{\prime} (p^2/M^2) ,
\end{split}\label{AdjointCorrelators}\eeq
we find
\beq
X_P = \frac{1}{\sqrt{2}} (e^{-i\theta}X + e^{i\theta} X^\dagger), \qquad X_M = -\frac{i}{\sqrt{2}} (e^{-i\theta} X - e^{i\theta}X^\dagger), 
\eeq
are two separate propagating degrees of freedom with masses $m_P, m_M$ respectively, where the angle $\theta$ is such that they are independent, and can be given by
\beq
e^{4i\theta} = \frac{\tilde{F}_0^{\prime\dagger} (0)}{\tilde{F}_0^{\prime} (0)}.
\label{DEFTheta}\eeq
Note that in many cases we find that $\tilde{F}_0^{\prime} (0)$ is real, in which case $X_P,X_M$ become the real and imaginary parts of $X$. In addition we define the correlators
\beq
\tilde{J}_0 (p^2/M^2) \equiv \bra J(p) F_2 (-p) \ket, \qquad  \tilde{J}_0^{\prime} (p^2/M^2) \equiv \bra J(p) F_2^{\dagger} (-p) \ket,
\eeq
we find
\beq\begin{split}
\tilde{C}_0 (p^2/M^2) \begin{array}{c}\\\sim \\  p^2 \rightarrow 0\end{array} & \frac{1}{p^2 + m_P^2}|\lambda_X|^2| e^{i\theta}\tilde{J}_0 (0) + e^{-i\theta}\tilde{J}_0^{\prime} (0) |^2 \\&+ \frac{1}{p^2 + m_M^2}|\lambda_X|^2| e^{i\theta}\tilde{J}_0 (0) - e^{-i\theta}\tilde{J}_0^{\prime} (0)|^2 ,
\end{split}\eeq
and thus we do not need to resum the $\tilde{C}_0$ contribution either. 
We can determine the masses to lowest order in $g\lambda_X$ as
\beq\begin{split}
m_P^2 =& \lambda_X^2\frac{M^2}{2} \bigg( \tilde{F}_0 (0) + |\tilde{F}_0^{\prime} (0)|\bigg) \\
m_M^2 =& \lambda_X^2\frac{M^2}{2} \bigg( \tilde{F}_0 (0) - |\tilde{F}_0^{\prime}(0)|\bigg).
\label{AdjointMasses}\end{split}\eeq

We can use the above to calculate the leading contribution to the sfermion masses due to the masses generated for the gauginos and adjoint scalars. Denoting the scale of supersymmetry breaking in the hidden sector as $\sqrt{\Delta}$ (where $\Delta$ can be taken to be $F, D$ for $F$ or $D$ term breaking respectively, or the square of the vacuum energy for a model-independent definition) we can compute the leading contribution in $\Delta/M^2$ at order $g^4 \lambda_X^2$ as follows. First consider that supersymmetry is restored above the scale $\sqrt{\Delta}$ and thus we can consider an expansion at low energies where 
\beq\begin{split}
\delta A_r =  - \lambda_X^2 \int \frac{d^4 p}{(2\pi)^4}\frac{1}{p^2} \bigg[& - \frac{4|H_{1/2}(0)|^2}{p^2 + m_D^2} + \frac{| e^{i\theta} \tilde{J}_0 (0) +  e^{-i\theta}\tilde{J}_0^{\prime} (0) |^2}{p^2+m_P^2} \\
&+ \frac{|  e^{i\theta}\tilde{J}_0 (0) -  e^{-i\theta}\tilde{J}_0^{\prime} (0) |^2}{p^2+m_M^2} \bigg].
\end{split}\eeq
This appears to have ultra-violet divergences. 
However,  these generate terms in the effective Lagrangian
\beq
\delta \C{L} \supset D\bigg( X \tilde{J}_0 (0) + X^\dagger \tilde{J}_0^\dagger (0) \bigg) - M H_{1/2}(0) \lambda\chi - M H_{1/2}^\dagger(0) \ov{\lambda}\ov{\chi }
\eeq
which at $\C{O}(\Delta/M^2)$ can only come from an effective term
\beq
\delta \C{L} \supset \int d^2 \theta W^{\prime}_\alpha W^\alpha X,
\eeq
where $W^{\prime}_\alpha$ is a spurion developing a vev $\theta_\alpha D$, and so
\beq\begin{split}
H_{1/2}(p^2) - \tilde{J}_0 (p^2) =& \C{O}(\Delta^2/M^4) \\
H_{1/2}^\dagger(p^2) - \tilde{J}_0^{\prime} (p^2) =& \C{O}(\Delta^2/M^4).
\label{Difference_Order_Two}\end{split}\eeq
We therefore find 
\beq
g_r^2 \delta A_r = 4 |m_D|^2 \log \frac{m_P^2}{|m_D|^2} + (m_D - m_D^\dagger)^2 \log \frac{m_P^2}{m_M^2} 
\label{General_Sfermions}\eeq
to first order in $\Delta/M^2$. If $H_{1/2}(0)$ is real, then we find
\beq
g_r^2 \delta A_r = 4 m_D^2 \log \frac{m_P^2}{m_D^2} 
\eeq
giving
\beq
\delta m_{\tilde{f}}^2 = \sum_{r=1}^3 C_2 (f; r) \frac{g_r^2 (m_D^{(r)})^2}{4\pi^2} \log \left(\frac{m_P^{(r)}}{m_D^{(r)}}\right)^2 ,
\label{Sfermions}\eeq
in agreement with the expressions in \cite{Fox:2002bu}.
Since we expect $\Delta/M^2 \ll g^2 \sim \lambda^2$, this becomes the dominant component if the two-loop mass squareds are $\C{O}(\Delta^4/M^6)$.

\section{Applications}
\label{APPLICATIONS}

In this section we shall show how to use the formalism developed above to calculate the gaugino and scalar masses in explicit examples. We will in turn illustrate the simple cases of supersymmetry breaking by D terms, R-symmetry preserving F-terms, and finally generic D and F terms.

\subsection{Example with D Term Breaking}

We consider as a first example the case of D-term supersymmetry breaking with messenger fields $Q,\tilde{Q}$ of mass $M$ with opposite charges under a hidden U(1) gauge field that develops a vev $W^{\prime}_{\alpha} = \theta_\alpha D$, coupled to an adjoint field with 
\beq
\C{J}_2 = Q\tilde{Q}. 
\eeq
This has been well studied in \cite{Polchinski:1982an,Fox:2002bu,Antoniadis:2006uj}. We find 
\beq
H_{1/2} (p^2/M^2) = \sqrt{2} \int \frac{d^4 q}{(2\pi)^4} \bigg( \frac{1}{q^2 + M_-^2} - \frac{1}{q^2 + M_{+}^2} \bigg) \frac{1}{(p+q)^2 +M^2}
\eeq
where $M_{\pm}^2 = M^2 \pm D$, and thus
\beq
m_D = g \lambda_X M H_{1/2} (0) = \frac{1}{\sqrt{2}} g \lambda_X \C{I} (M, D) 
\label{DTermG}\eeq
where we have defined
\beq\begin{split}
\C{I} (M, \Delta) &\equiv 2\frac{1}{(4\pi)^2} \frac{\Delta}{M}  \bigg[ \frac{(1-\Delta/M^2) \log (1-\Delta/M^2) + (1+\Delta/M^2)\log (1+\Delta/M^2)}{\Delta^2/M^4} \bigg] \\
&\sim \frac{2}{(4\pi)^2} \frac{\Delta}{M} + ...
\end{split}\label{DefI}\eeq
Since the gaugino masses in this model are pure Dirac, we find that for degenerate messenger masses $M$ under $SU(3), SU(2)$ and $U(1)_1$ that the ratio of gaugino masses is
\beq
m_1 : m_2 : m_3 = \lambda_1\sqrt{\alpha_1 } : \lambda_2 \sqrt{\alpha_2 } : \lambda_3\sqrt{\alpha_3 }
\eeq
The parameters $\lambda_i$ are not independent, however; they are determined by running the coupling from the unification scale.

\subsection{Example with R-symmetric F Term}

As another example, consider the model of \cite{Amigo:2008rc}: the standard model gauge group is embedded in an $SU(5) \subset SU(6)$, and there is a hidden $U(1)$ group. The field content is given by an $SU(6)$ adjoint $M$, $SU(5)$ adjoints $\Phi, M'$; fields $N$ and $\varphi$ in the fundamental of $SU(5)$ and $\bar{N}, \bar{\varphi}$ in the antifundamental, supplemented by a singlet pair $\psi, \bar{\psi}$. The superpotential is
\beq
W = W_1 + \lambda(\bar{\varphi} M \varphi + \kappa' \bar{\psi} X \psi + \kappa \bar{\varphi} N \psi + \kappa \bar{\psi} \bar{N} \varphi) - f^2(X+\omega \mathrm{Tr} M),
\eeq
where
\beq
W_1 = y(\bar{\varphi}\Phi N - \bar{N} \Phi \varphi).
\eeq
$\bar{\psi} \psi$ develops a vev equal to $\Mm^2=|v|^2=f^2/\lambda \kappa'$, and $\bra F_{\mathrm{Tr} M}\ket = \omega f^2$. The $\varphi, \bar{\varphi}^*$ scalars mix into $\phi_{\pm} = \frac{1}{\sqrt{2}} ( \bar{\varphi}^* \pm \varphi)$ with masses $m_{\pm}^2 = (1\pm z) \Mm^2$, where $z= \omega \kappa'/\kappa^2$. The scalars $N, \bar{N}$ acquire masses $\Mm$, while there is a Dirac mass term $v e^{i\xi/v} \varphi \bar{N} + v e^{-i\xi/v} \bar{\varphi} N$ (where we have used the same symbol for superfield, scalar \emph{and} fermion). The currents are 
\begin{align}
j_{\alpha} &=-\sqrt{2}i \bigg[\varphi^* \varphi_{\alpha} - \bar{\varphi}^* \bar{\varphi}_{\alpha} + N^* N_{\alpha} - \bar{N}^* \bar{N}_\alpha + M^* M_{\alpha} + M^{\prime *} M^{\prime}_{\alpha} + \Phi^* \Phi_{\alpha} \bigg]\nonumber \\
j_{2\alpha} &=(N \bar{\varphi}_\alpha + \bar{\varphi} N_{\alpha} - \bar{N} \varphi_{\alpha} - \varphi \bar{N}_{\alpha}) \nonumber \\
J &= \varphi^* \varphi - \bar{\varphi}^* \bar{\varphi} + N^* N - \bar{N}^* \bar{N} + M^* M + M^{\prime *} M^{\prime} + \Phi^* \Phi \nonumber \\
F_2 &= N F_{\bar{\varphi}} + F_N \bar{\varphi} - \bar{N} F_{\varphi} - F_{\bar{N}} \varphi 
\end{align}
and $j_{\mu}$ defined similarly. We have also suppressed gauge indices.

This model contains only Dirac gaugino masses; thus the gaugino masses are given by equation (\ref{DiracOnly}). We find
\beq\begin{split}
M H_{1/2}(p^2/M^2) = & \frac{1}{\sqrt{2}}\Mm \cos \xi/v \int \frac{d^4 q}{(2\pi)^4} \frac{1}{(q+p)^2 + \Mm^2} \\ 
& \times \bigg( \frac{1}{q^2 + m_+^2} + \frac{1}{q^2 + m_-^2} - \frac{2}{q^2 + \Mm^2}\bigg) 
\end{split}\eeq
in agreement with \cite{Amigo:2008rc}.

At order $g^4$, the sfermion masses are given by the contributions of the fields $\varphi, \bar{\varphi},N,\bar{N}$ only. We shall compute these using current correlators and equation (\ref{MSSsfermion}). We begin with $\tilde{C}_1$, given by
\beq
\tilde{C}_1 = -\frac{1}{3p^2} ( \eta^{\mu \nu} - 4 \frac{p^\mu p^\nu}{p^2}) \bra j_{\mu} (p) j_{\nu} (-p) \ket,
\eeq
since there are contact terms
\beq
\bra j_{\mu} (p) j_{\nu} (-p) \ket + a \eta_{\mu \nu}= -(p^2 \eta_{\mu \nu} - p_\mu p_\nu ) \tilde{C}_1.
\eeq
We then require the two point current correlator: 
\beq\begin{split}
\bra j_\mu (x) j_{\nu} (0) \ket =& 2 \bigg[ ( \partial_{\mu} D (x; m_+) \partial_\nu D (x; m_+) - D(x; m_+) \partial_\mu \partial_\nu D (x; m_+) \\
&+ m_+ \leftrightarrow m_- \\ 
& + 2  \partial_{\mu} D (x; \Mm ) \partial_\nu D (x; \Mm) - 2 D(x; \Mm) \partial_\mu \partial_\nu D (x; \Mm) \\
&+ 4 \eta_{\mu \nu} ( \partial^{\rho} D(x;\Mm) \partial_{\rho} D(x; \Mm) - \Mm^2  (D(x;\Mm))^2 ) \\
&- 8 \partial_\mu D(x;\Mm) \partial_{\nu} D(x;\Mm) \bigg]
\end{split}\eeq
which leads to
\beq\begin{split}
\tilde{C}_1 =& -\frac{2}{3p^2} \int \frac{d^4 q}{(2\pi)^4} \bigg( \frac{ (p+q) \cdot (p + 2q)}{(q^2 + m_{+}^2)((p+q)^2 + m_{+}^2)} + m_+ \leftrightarrow m_-  \\
& + \frac{2 p^2 + 14 p \cdot q + 12q^2 + 16\Mm^2}{(q^2+\Mm^2)((p+q)^2 + \Mm^2)} \\
& -\frac{4}{q^2 + m_+^2} - \frac{4}{q^2 + m_-^2} - \frac{8}{q^2 + \Mm^2}\bigg).
\end{split} 
\eeq

We find, using the notation of \cite{Martin:1996zb},
\begin{align}
-3\tilde{C}_{1} \rightarrow& -  \bra 0 | m_+ | m_+ \ket - 4m_+^2 \bra 0,0 | m_+ | m_+ \ket  + m_+ \leftrightarrow m_- \nonumber \\
&  - 10 \bra 0 | \Mm | \Mm \ket + 8 \Mm^2 \bra 0,0 | \Mm | \Mm \ket \nonumber \\
&- 4 \bra 0,0 \ket \bra m_+ \ket - 4 \bra 0,0 \ket \bra m_- \ket + 8 \bra 0,0 \ket \bra \Mm \ket
\end{align}
and, similarly
\begin{align}
-\tilde{C}_0 \rightarrow& -2 \bra 0 | m_+ | m_- \ket - 2 \bra 0 | \Mm | \Mm \ket \nonumber \\
4 \tilde{C}_{1/2} \rightarrow& -8 \bra 0,0 \ket \bra \Mm \ket + 4 \bra 0,0 | m_+ \ket +4 \bra 0,0 | m_- \ket  +4 \bra 0 | m_+ | \Mm \ket  \nonumber \\
& + 4 z \Mm^2 \bra 0,0 | m_+ | \Mm \ket +4 \bra 0 | m_- | \Mm \ket -4z \Mm^2 \bra 0,0 | m_- | \Mm \ket  \nonumber \\
& +8 \bra 0 | \Mm | \Mm \ket 
\end{align}
which, when integrated using the expressions of \cite{Martin:1996zb}, yields
\beq\begin{split}
A_r =& \frac{\Mm}{(4\pi)^4} \bigg( 2 (1 + \log (1-z^2))( (1+z)\log (1+z) + (1-z)\log(1-z)) \\
&- 4z \li_2 (z) + 2z \li_2 (z/(z-1)) - (1+z) \li_2 (2z/(z-1))  \\
&+ (2+z) \li_2 (z^2) - 2z \li_2 (z/(1+z)) -(1-z) \li_2 (2z/(1+z)) \bigg).
\end{split}\eeq
After some manipulations one can show that
\beq\begin{split}
A_r =& \frac{2\Mm}{(4\pi)^4} (1+z)\bigg[\log(1+z) - 2 \li_2 (z/(1+z)) + \frac{1}{2} \li_2 (2z/(1+z)) \bigg] + (z \leftrightarrow -z).
\end{split}
\eeq
in agreement with \cite{Amigo:2008rc}.

\subsection{D and F Term Breaking}
\label{DandF}

Suppose now that we return to the case of a single messenger pair, but consider both D and F term breaking. The mass eigenstates of the scalars are given by
\beq\begin{split}
\phi_+ &= \frac{1}{\sqrt{|F|^2 + |D+\Delta|^2}} ( (D+\Delta)Q + F^{\dagger} \tilde{Q}^\dagger) \\
\phi_- &= \frac{1}{\sqrt{|F|^2 + |D-\Delta|^2}} ( (D-\Delta)Q^\dagger + F \tilde{Q})
\end{split}\eeq
with masses $m_{\pm}^2 = \Mm^2 \pm \Delta$ where $\Delta = \sqrt{D^2 + |F|^2}$. This leads to two-point functions
\beq\begin{split}
\bra Q(x) Q^\dagger (0) \ket &= \frac{D+\Delta}{2\Delta} D(x;m_+) + \frac{\Delta-D}{2\Delta} D(x;m_-) \\
\bra \tilde{Q}^\dagger(x) \tilde{Q} (0) \ket &= \frac{\Delta-D}{2\Delta} D(x;m_+) + \frac{\Delta+D}{2\Delta} D(x;m_-) \\
\bra Q(x) \tilde{Q} (0) \ket &= \frac{F^{\dagger} }{2\Delta} D(x;m_+) - \frac{F^{\dagger}}{2\Delta} D(x;m_-)
\end{split}\eeq
leading to 
\begin{align}
H_{1/2}(p^2/\Mm^2) &= \frac{1}{\sqrt{2}}\frac{D}{\Delta} \C{I}(M, \Delta) \\
B_{1/2} (p^2/\Mm^2) &= \frac{F}{\Delta} \C{I}(M, \Delta)\\
I_{1/2} (p^2/\Mm^2) &= \frac{1}{2} \frac{F^{\dagger}}{\Delta} \C{I}(M, \Delta)
\end{align}
where $\C{I}(M, \Delta)$ was defined in equation (\ref{DefI}). 
Then we calculate the gaugino masses to be
\beq
m_{\pm} = \C{I}(M, \Delta)  \frac{1}{2\Delta} \bigg[ \frac{1}{2} \lambda_X^2 F^\dagger + g^2 F \pm \sqrt{ (\frac{1}{2} \lambda_X^2 F^\dagger - g^2  F)^2 + 2 g^2 \lambda_X^2 D^2} \bigg] . 
\eeq

\section{Adjoint Scalar Masses}
\label{SCALARS}

In this section we consider the generation of masses for the adjoint scalars. Recall that the adjoints can be decomposed into two components $X_R, X_I$ with masses $m_R, m_I$ that can be calculated using current correlators using equation (\ref{AdjointMasses}). From that we see that if
\beq
|\tilde{F}_0^{\prime}(0)| > |\tilde{F}_0 (0)| 
\eeq
then we have a tachyonic spectrum. In this section we calculate the masses for several models. The simpler models yield tachyons, but we calculate the masses in general for models with F-terms, and those with D-terms, and show how tachyon-free models can be found in each.

\subsection{One Messenger Pair}

Here we consider the case of one pair of $SU(5)$ messengers as in section \ref{DandF}. To compute the scalar mass correction, we have two options. One is to treat the auxiliary fields as dynamical, and construct propagators for them and their mixing with the fields $Q, \tilde{Q}$. The unitary transformation that rotates $Q, \tilde{Q}$ can also rotate their auxiliary fields $F_Q, \tilde{F}_Q$ as
\beq
\left( \begin{array}{c} F_Q \\ \tilde{F}_Q^{\dagger} \end{array}\right) = \frac{1}{\sqrt{2\Delta}}\left( \begin{array}{c} \frac{F}{\sqrt{\Delta+D}} F_+ + \frac{F}{\sqrt{\Delta-D}} F_- \\ \sqrt{\Delta+D} F_+ - \sqrt{\Delta-D} F_-  \end{array}\right) 
\eeq
giving propagators
\beq\begin{split}
\bra \phi_+ (x) F_+^\dagger (0) \ket &= -\Mm D(x; m_+)\\
\bra F_+ (x) F_+^{\dagger} (0) \ket &= (\Box +\Delta)D (x; m_+) \\
\bra \phi_- (x) F_-^\dagger (0) \ket &= -\Mm D(x; m_-)\\
\bra F_- (x) F_-^{\dagger} (0) \ket &= (\Box - \Delta) D (x; m_-)
\end{split}\eeq
and thus we find
\beq
F_2 = \frac{|F|}{2\Delta} ( \hat{F}_2^{M\dagger} -\hat{F}_2^M)  + \frac{(\Delta +D)}{2\Delta}\hat{F}_2^H + \frac{(\Delta - D)}{2\Delta}\hat{F}_2^{H\dagger} + q \tilde{q}
\eeq
where
\begin{align}
\hat{F}_2^M &= \phi_+ F_-^{\dagger} + \phi_- F_+^\dagger  \\
\hat{F}_2^H &= \phi_+ F_+^{\dagger} + \phi_-^{\dagger} F_- .
\end{align}
 In this case, the $\C{O}(\lambda^2)$ mass for the scalar is given by $\bra F_2 (p) F^{\dagger}_2(-p) \ket$, but also the correlator $\bra F_2 (p) F_2(-p) \ket$ gives the same scalar contribution to $XX + \ov{X}\ov{X}$; the only difference comes from the fermions. We thus compute \footnote{Of course, it is easier to compute the contributions by considering the couplings in the Lagrangian
\beq
\C{L} \supset - \lambda_X^2 [ Q^\dagger X \ov{X} Q + \tilde{Q} \ov{X} X \tilde{Q} ^\dagger ] - \lambda_X \Mm \{[ Q^\dagger (X + \ov{X}) Q + \tilde{Q} (X+ \ov{X}) \tilde{Q}^\dagger] + X q \tilde{q} + \ov{X} \bar{q} \bar{\tilde{q}}\}
\eeq
which is left invariant by the diagonalisation of $Q, \tilde{Q}$. This calculation is to illustrate the method.}
\begin{align}
\bra \hat{F}_2^H (p) \hat{F}_2^{H\dagger} (-p) \ket =& \bigg[ \bra m_+ \ket + \bra m_- \ket - \bra m_+ |m_+\ket - \bra m_-|m_- \ket \bigg] \nonumber \\
\bra \hat{F}_2^H (p) \hat{F}_2^H (-p) \ket =& -\Mm^2 \bigg[ \bra m_+ |m_+\ket + \bra m_-|m_- \ket \bigg] \nonumber\\
\bra \hat{F}_2^M (p) \hat{F}_2^{M\dagger} (-p) \ket =&  \bigg[ \bra m_+ \ket + \bra m_- \ket -2\Mm^2 \bra m_+ |m_-\ket \bigg] \nonumber \\
\bra \hat{F}_2^M (p) \hat{F}_2^M (-p) \ket =& -2 \Mm^2 \bra m_+ | m_- \ket \\
\bra q \tilde{q} (p) q \tilde{q} (-p) \ket =& 2\Mm^2 \bra \Mm | \Mm \ket \\
\bra q \tilde{q} (p) q^\dagger \tilde{q}^\dagger (-p) \ket =& (2 \Mm^2 + p^2) \bra \Mm | \Mm \ket - 2 \bra \Mm \ket 
\end{align}

Thus we find 
\beq\begin{split}
M^2\tilde{F}_0 (p^2/M^2) =& \bigg[ \bra m_+ \ket + \bra m_- \ket - 2 \bra \Mm \ket \bigg] + (2 \Mm^2 + p^2) \bra \Mm | \Mm \ket \\
&- \Mm^2 \bigg( \bra m_+ | m_+ \ket +  \bra m_- | m_- \ket \bigg)\\
M^2\tilde{F}_0^\prime (p^2/M^2) =& \Mm^2  \bigg[ 2 \bra \Mm | \Mm \ket - \bra m_+ | m_+ \ket - \bra m_- | m_- \ket \bigg].
\end{split}\eeq
It is straightforward to see that
\beq\begin{split}
M^2 \tilde{F}_0 (0) &= \frac{\Mm^2}{(4\pi)^2} \bigg[ (2+z)\log (1+z) + (2-z) \log (1-z) \bigg] = -\frac{ \Mm^2}{(4\pi)^2} ( \frac{1}{3}z^4 + ...)\\
M^2 \tilde{F}_0^\prime (0) &= \frac{\Mm^2}{(4\pi)^2} \log (1-z^2) = -\frac{z^2 \Mm^2}{(4\pi)^2} ( 1 + \frac{1}{2}z^2  + ...),
\end{split}\eeq
and therefore there is a tachyon.

We can also compute the correlators $\tilde{J}_0 (p^2/M^2), \tilde{J}_0^\prime (p^2/M^2)$; we find
\beq
J = Q^* Q - \tilde{Q}^* \tilde{Q} = \frac{1}{\Delta} \bigg[ D (\phi_+^\dagger \phi_+ - \phi_-^\dagger \phi_-) - |F| (\phi_+^\dagger \phi_- + \phi_-^\dagger \phi_+) \bigg]
\eeq
which leads us to
\beq\begin{split}
\tilde{J}_0 (0)&= \frac{\Mm D}{\Delta} \bigg[ \bra m_+ | m_+ \ket - \bra m_- | m_- \ket \bigg]\\
&=\frac{\Mm D}{\Delta} \log (m_+^2/m_-^2) \\
&=\frac{1}{(4\pi^2)}\frac{2D}{\Mm} + ...
\end{split}\eeq
which is equal to $H_{1/2}(0)$ to leading order in $\Delta$, compatible with equation (\ref{Difference_Order_Two}). We also find that $\tilde{J}_0^\prime =\tilde{J}_0$ here.

\subsection{Multiple Messengers}

\subsubsection{F Terms}

We begin this subsection by analysing again the model of \cite{Amigo:2008rc}. Consider the interaction with the adjoint scalar in the Lagrangian with the scalar vev:
\beq\begin{split}
\C{L} \supset &y^2 |\Phi|^2 \bigg[ |\varphi|^2 + |\bar{\varphi}|^2 + |N|^2 + |\bar{N}|^2 \bigg] \\
& +(\Phi + \Phi^*) \Mm y \bigg[ - |\varphi|^2 + |\bar{\varphi}|^2 + |N|^2 - |\bar{N}|^2 \bigg] \\
& +y \Phi (\bar{\varphi}_\alpha N^\alpha - \bar{N}_\alpha \varphi) + c.c
\end{split}\eeq

Note that we can write $|\bar{\varphi}|^2 - |\varphi|^2 = \varphi_+^* \varphi_- + \varphi_-^* \varphi_+$. This leads to 
\beq\begin{split}
M^2\tilde{F}_0 (p^2/M^2) =& -\Mm^2 \bigg[ 2 \bra m_+ | m_- \ket + 2 \bra \Mm | \Mm \ket \bigg] \\
&+  4( \Mm^2 + p^2/2 ) \bra \Mm | \Mm\ket + \bra m_+ \ket + \bra m_- \ket -2\bra \Mm \ket \\
M^2\tilde{F}_0^\prime (p^2/M^2) =& -\Mm^2 \bigg[ 2 \bra m_+ | m_- \ket + 2 \bra \Mm | \Mm \ket \bigg] + 4 \Mm^2 \bra \Mm | \Mm \ket,
\end{split}\eeq
and thus
\beq\begin{split}
M^2 \tilde{F}_0 (0) =& -\frac{\Mm^2}{16\pi^2}  \bigg[ 2 + \frac{(1-z)^2}{z} \log (1-z) - \frac{(1+z)^2}{z} \log (1+z) \bigg] \\
&\sim \frac{\Mm^2}{16\pi^2}\bigg[\frac{2}{3}z^2 + \frac{1}{15}z^4  + ...\bigg]  \\
M^2\tilde{F}_0^\prime (0) =& -\frac{\Mm^2}{16\pi^2} \frac{1}{z} \bigg[ 2z - (1+z) \log (1+z) + (1-z) \log (1-z) \bigg] \\
&\sim -\frac{\Mm^2}{16\pi^2} \bigg[ \frac{1}{3} z^2 + \frac{1}{10}z^4  + ... \bigg].
\end{split}\eeq
Hence the two states of the adjoint have masses $z^2y^2 \Mm^2/32\pi^2, z^2y^2\Mm^2/96\pi^2$, so there is no tachyon. This shows that we may avoid a tachyon by using multiple messenger fields. 

Consider that for $F$-term supersymmetry breaking a spurion $\Xi$ aquires a vev $\theta^2 F$; we must generate a term in the K\"ahler potential of
\beq
\Delta K \supset \frac{\Xi \Xi^\dagger}{\Lambda^2} \tr(\C{X} \C{X}^\dagger) .
\eeq
We wish to find explicit models that will generate such terms. To do this we can either compute diagrams, or use the result of \cite{deWit:1996kc,Pickering:1996gt,Grisaru:1996ve,Brignole:2000kg}: the correction to the K\"ahler potential at one loop, ignoring the gauge-dependent contribution (since we are considering spurions which are singlets of the standard model gauge groups) is
\beq
\Delta K = - \frac{1}{32\pi^2}  \tr \bigg( \C{M}_\phi^2 \bigg(  \log \frac{\C{M}_\phi^2}{\Lambda^2}-1\bigg)\bigg) 
\eeq
where for canonical tree-level K\"ahler potential, and generic superpotential $W$
\beq
\C{M}_\phi^2 = \ov{W}_{\bar{i}\bar{k}} \delta^{\bar{k}k} W_{kj}.
\eeq
We can now use this expression and expand to quartic order in the fields $\C{X},\Xi$. For example, the simplest case that we can consider is the superpotential
\beq
W = Z \tr (Q \tilde{Q}) + Q \C{X} \tilde{Q}
\eeq
where $Z = M + \Xi $, generates 
\beq
\Delta K = -\frac{1}{16\pi^2} \tr \bigg( (Z^\dagger + \C{X}^\dagger)(Z+ \C{X}) \bigg( \log \frac{(Z^\dagger + \C{X}^\dagger)(Z+ \C{X})}{\Lambda^2} -1 \bigg) \bigg) 
\eeq
which, when expanded to quartic order and we consider terms proportional to $\Xi^\dagger \Xi$ gives (when we consider that $\ov{Z}, Z$ should be multiplied by the unit matrix for the gauge indices, and thus commute with $\ov{\C{X}},\C{X}$) 
\beq
\Delta K \supset \frac{1}{32\pi^2} \frac{\Xi^\dagger \Xi}{M^2}  \tr ( \ov{\C{X}}^2 + \C{X}^2 )
\eeq
as expected - this gives a tachyonic mass. However, if we now generalise the model a little we find that we can generate precisely the term required. Consider a set of $N_{\mathrm{mess}}$ pairs of fundamental and antifundamental fields $Q_i, \tilde{Q}_{\bar{i}}$ of equal masses $\Mm$. We can write
\beq
W = \Mm \tr(Q_i \tilde{Q}_{\bar{i}}) \delta_{i \bar{i}} + \lambda_{i \bar{j}} Q_i \C{X} \tilde{Q}_{\bar{j}} + \Xi \mu_{i \bar{j}} \tr (Q_i \tilde{Q}_{\bar{j}})
\eeq
and then we find, after some algebra, that (for simplicity taking $\lambda, \mu$ to be real)
\beq\begin{split}
\Delta K \supset \frac{1}{16\pi^2} \frac{\Xi^\dagger \Xi}{\Mm^2} \frac{1}{6} \bigg[  \tr (\C{X}^\dagger \C{X}) \tr \bigg(  [\lambda, \mu][\lambda^T,\mu^T] - 2[\lambda, \lambda^T][\mu ,\mu^T] \bigg) \\
+   \tr(\ov{\C{X}}^2 + \C{X}^2 )  \tr \bigg( \mu^T (\mu \lambda^2 + \lambda \mu \lambda + \lambda^2 \mu)\bigg) \bigg]
\end{split}\eeq
and so to ensure the absence of tachyons we require
\beq\begin{split}
\tr \bigg( 2[\lambda, \lambda^T][\mu ,\mu^T]- [\lambda, \mu][\lambda^T,\mu^T] \bigg) &> 2 \bigg|\tr \bigg( \mu^T (\mu \lambda^2 + \lambda \mu \lambda + \lambda^2 \mu)\bigg)\bigg|.
\end{split}\eeq
We also find a linear term proportional to $ \Xi \Xi^\dagger \tr(\{\mu, \mu^\dagger\} (\lambda \C{X} + \lambda^\dagger \C{X}^\dagger))$, which must vanish to prevent a vev for $X, X^\dagger$. These constraints can be satisfied for example with $N_{\mathrm{mess}}=2$ and the following choices
\beq
\lambda = y\2b2[1,0][0,-1], \qquad \mu = \2b2[0,1][0,0]
\eeq
which gives
\beq\begin{split}
\Delta K &\supset -\frac{y^2}{16\pi^2} \frac{\Xi^\dagger \Xi}{\Mm^2} \frac{1}{6} \bigg[ 4 tr (\C{X}^\dagger \C{X}) - tr (\ov{\C{X}}^2 + \C{X}^2 ) \bigg] \\
&\supset -\frac{y^2}{96\pi^2} \frac{\Xi^\dagger \Xi}{\Mm^2} \bigg[ \tr(\frac{1}{2} | \C{X} + \ov{\C{X}}|^2 + \frac{3}{2} |\C{X} - \ov{\C{X}}|^2 )\bigg].
\end{split}\eeq

\subsubsection{D Terms}

Now we consider stabilising models with a D-term. In this case, we do not calculate the K\"ahler potential, but rather the Coleman-Weinberg potential:
\beq
V= \frac{1}{64\pi^2} \mathrm{Str} (\C{M}^4 \log \C{M}^2)
\eeq
We can consider a similar model to the above, with 
\beq
W = \Mm tr(Q_i \tilde{Q}_{\bar{i}}) \delta_{i \bar{i}} + \lambda_{i \bar{j}} Q_i \C{X} \tilde{Q}_{\bar{j}} 
\eeq
but now the supersymmetry is broken by a term
\beq
\Delta V = D \bigg( \sum_i e_i |Q_i|^2 + \tilde{e}_i |\tilde{Q}_i|^2 \bigg)
\eeq
where the $e_i, \tilde{e}_i$ are charges under a hidden gauge group whose coupling has been absorbed into $D$. We can thus express the mass matrix as
\beq
\C{M}^2 = M_{ij}^2 \oplus \tilde{M}_{\bar{i} \bar{j}}^2
\eeq
where
\beq\begin{split}
M_{ij}^2 &= \Mm^2 [( 1+ \lambda X/\Mm)(1+ \lambda^{\dagger} X^{\dagger})/\Mm]_{ij} + D e_i \delta_{ij} \\
\tilde{M}_{\bar{i} \bar{j}}^2 &= \Mm^2 [( 1+ \lambda^T X^T/\Mm)(1+ \lambda^{*} X^{*}/\Mm)]_{\bar{i} \bar{j}} + D \tilde{e}_{\bar{i}} \delta_{\bar{i}\bar{j}}
\end{split}\eeq
where the trace over gauge indices has been suppressed. If we make the further simplifying assumption that $\tilde{e}_{\bar{i}} = - e_i$, then writing $\hat{e} \equiv e_i \delta_{ij}$ we find that to quartic order in $D/\Mm^2$ 
\beq\begin{split}
V_1 =& \frac{\Mm^2}{32\pi^2} \tr \bigg[ 2\Mm(X \lambda + X^\dagger \lambda^\dagger) \bigg((1+\frac{ \hat{e} D}{\Mm^2})  \log (1+ \frac{\hat{e}D}{\Mm^2}) \\
& \qquad\qquad+(1- \frac{ \hat{e} D}{\Mm^2})  \log (1- \frac{\hat{e}D}{\Mm^2})\bigg)\\
&+ XX^\dagger \bigg( \frac{3}{2} \frac{D}{\Mm^2} (\hat{e}[\lambda,\lambda^\dagger]) - \frac{1}{3} \frac{D^3}{\Mm^6} (\hat{e}^3 [\lambda,\lambda^\dagger])\bigg)\\
&+ XX^\dagger \bigg( \frac{D^2}{\Mm^4} \bigg\{ \frac{1}{3}([\hat{e},\lambda][\lambda^\dagger,\hat{e}])\bigg\} \\
&+\frac{D^4}{\Mm^8}\bigg\{\frac{2}{15} (\hat{e}^4 \lambda \lambda^\dagger) - \frac{1}{5} (\hat{e}^4 \lambda^\dagger\lambda) - \frac{1}{5} ( \lambda^\dagger\hat{e}\lambda\hat{e}^3) - \frac{1}{5} ( \lambda\hat{e}\lambda^\dagger\hat{e}^3) -  \frac{1}{5} ( \lambda\hat{e}^2\lambda^\dagger\hat{e}^2) \bigg\}\bigg)\\
& - XX \bigg( \frac{D^2}{\Mm^4} \bigg\{ \frac{2}{3}(\lambda^2 \hat{e}^2) + \frac{1}{3} (\lambda \hat{e} \lambda \hat{e}) \bigg\} \\
&- \frac{D^4}{\Mm^8}\bigg\{ \frac{1}{5} (\lambda^2 \hat{e}^4) + \frac{1}{5} (\lambda \hat{e} \lambda \hat{e}^3) + \frac{1}{10} (\lambda \hat{e}^2 \lambda \hat{e}^2) \bigg\} \bigg) \\
& - X^\dagger X^\dagger \bigg( \frac{D^2}{\Mm^4} \bigg\{ \frac{2}{3}((\lambda^\dagger)^2 \hat{e}^2) + \frac{1}{3} (\lambda^\dagger \hat{e} \lambda^\dagger \hat{e}) \bigg\} \\
&- \frac{D^4}{\Mm^8}\bigg\{ \frac{1}{5} ((\lambda^\dagger)^2 \hat{e}^4) + \frac{1}{5} (\lambda^\dagger \hat{e} \lambda^\dagger \hat{e}^3) + \frac{1}{10} (\lambda^\dagger \hat{e}^2 \lambda^\dagger \hat{e}^2) \bigg\} \bigg) \bigg]
\end{split}\eeq
We see that we can ensure the absence of a linear term for a $U(1)$ adjoint by requiring $\tr(\lambda \hat{e}^{2n}) = 0$; this can be achieved by choosing charges $0,\pm 1$ and $\tr'(\lambda)=0$, where the prime denotes that the trace should be taken over indices for which $e_i \ne 0$. We can also read off the condition to ensure no tachyons at small $D$ as being $\tr(\hat{e}[\lambda,\lambda^\dagger]) > 0$ or $\tr(\hat{e}[\lambda,\lambda^\dagger]) = 0$,
\beq
\tr([\hat{e}, \lambda] ([\hat{e},\lambda])^\dagger) > |4\tr(\lambda^2 \hat{e}^2) + 2\tr(\lambda \hat{e} \lambda \hat{e})|.
\eeq  
We can solve this, for example, with the following choices:
\beq
\lambda = y\2b2[0,1][-1,0], \qquad \hat{e} = \2b2[1,0][0,-1]
\eeq
This corresponds to a model with fields $Q_1, Q_2, \tilde{Q}_1, \tilde{Q}_2$ with $U(1)$ charges $1,-1,-1,1$ and superpotential
\beq
W = \Mm Q_1 \tilde{Q}_1 + \Mm Q_2 \tilde{Q}_2 + y (Q_1 \C{X} \tilde{Q}_2 - Q_2 \C{X} \tilde{Q}_1).
\eeq
Here we can explicitly calculate the potential to be
\beq\begin{split}
V (\Sigma_R, \Sigma_I) = \frac{\Mm^4}{16\pi^2} \tr\bigg\{&\bigg(1 + \frac{1}{2} (\Sigma_R^2 + \Sigma_I^2) + \sqrt{g^2 D^2/\Mm^4 + 2 \Sigma_I^2}\bigg)^2 \\
&\times \log \bigg[1 + \frac{1}{2} (\Sigma_R^2 + \Sigma_I^2) + \sqrt{g^2 D^2/\Mm^4 + 2 \Sigma_I^2}\bigg] \\
&+\bigg(1 + \frac{1}{2} (\Sigma_R^2 + \Sigma_I^2) - \sqrt{g^2 D^2/\Mm^4 + 2 \Sigma_I^2}\bigg)^2 \\
&\times \log \bigg[1 + \frac{1}{2} (\Sigma_R^2 + \Sigma_I^2) - \sqrt{g^2 D^2/\Mm^4 + 2 \Sigma_I^2}\bigg]\bigg\}
\end{split}\eeq
for $\Sigma_R = \sqrt{2}y\Re(X)/\Mm, \Sigma_I = \sqrt{2}y \Im(X)/\Mm$. This is strictly positive, with a minimum at $\Sigma_R=\Sigma_I = 0$, for $D^2/\Mm^4 < 1$. Unfortunately this model does not generate gaugino masses at one loop. In general, to calculate the gaugino masses we write the currents 
\beq\begin{split}
j_\alpha =& \sqrt{2}\sum_i v_i Q_i^* Q_{i\alpha} + \sqrt{2}\sum_{\bar{i}}\tilde{v}_{\bar{i}} \tilde{Q}_{\bar{i}}^* \tilde{Q}_{\bar{i}\alpha} \\
j_{2\alpha} =& \lambda_{i \bar{j}} ( Q_i \tilde{Q}_{\bar{j}\alpha} + Q_{i \alpha} \tilde{Q}_{\bar{j}})
\end{split}\eeq
where $v_i, \tilde{v}_{\bar{i}}$ describe the ``charges'' of the messengers under the visible gauge group. For the $SU(N)$ gauge groups we must take $v_i = - \tilde{v}_{\bar{i}} = 1$, while for the hypercharge we can choose the charges but take $v_i = - \tilde{v}_{\bar{i}}$ for simplicity. Then defining $\hat{v}=\mathrm{diag}(v_i)$, 
we can calculate the Dirac gaugino masses to be (for $e_i = 0,\pm 1$)
\beq
m_D = \frac{1}{\sqrt{2}} g\tr (\lambda \hat{e} \hat{v}) \C{I}(\Mm, D)
\eeq
where $\C{I}(\Mm,\Delta)$ was defined in equation (\ref{DefI}). Now we can easily write down a model that avoids a tachyonic direction while allowing for gaugino masses:
\beq
\lambda = y\2b2[1,a][-a,-1], \qquad \hat{e} = \2b2[1,0][0,-1], \qquad \hat{v} = \2b2[1,0][0,1],
\eeq
where $2|a|^2 > |a^2-3|$. We then find scalar mass squareds of 
\beq
m_{P,M}^2= \frac{(2|a|^2 \pm |a^2-3|)|y|^2D^2}{24\Mm^2\pi^2}, 
\eeq
with $e^{4i\theta} = \frac{\bar{a}^2-3}{a^2-3}$ (c.f. equation (\ref{DEFTheta})) and gaugino masses of $\sqrt{2}gy \C{I}(\Mm, D)$. With regards to the visible sector sfermion mass squareds, the two loop contribution will be $\C{O}(D^4/\Mm^6)$ by the reasoning of \cite{Distler:2008bt,Intriligator:2008fr,Nakayama:2007cf}, and so provided $D^2/\Mm^4 < \lambda^2$ the dominant contribution comes at three loops. Calculated using equation (\ref{Sfermions}), this is
\beq\begin{split}
m_{\tilde{f}}^2 =& \sum_{r=1}^3 \frac{\alpha_r^2  C_2 (f;r)}{32\pi^4} \frac{D^2}{M^2} \bigg\{ 4|y|^2 \log \bigg[\frac{\pi}{3\alpha_r} (2|a|^2+|a^2-3|) \bigg] \\
& + (y-y^*)^2 \log \bigg[ \frac{2|a|^2+|a^2-3|}{2|a|^2-|a^2-3|} \bigg]\bigg\}
\end{split}\eeq

For $a,y$ real, we simply require to $a^2>1$, and then the mass squareds are 
\beq
m_{\tilde{f}}^2 = \sum_{r=1}^3 C_2 (f;r)  \frac{y^2 \alpha_r^2}{8\pi^4} \frac{D^2}{\Mm^2} \log \bigg[ \frac{\pi}{\alpha_r} (a^2 -1) \bigg].
\eeq
We thus have an R-symmetric model with D-term supersymmetry breaking that depends upon two real parameters $D, \Mm$ and two complex ones $a, y$.

\section{General Kinetic Mixing}
\label{MIXING}

The calculation of kinetic mixing between U(1) gauge fields \cite{Okun:1982xi,Polchinski:1982an,Holdom:1985ag,Dienes:1996zr,Feldman:2007wj,Ibarra:2008kn,Ahlers:2008qc} can also be recast into the evaluation of current correlators: consider initially  a single hidden U(1) gauge field with field strength $W^{\prime}_{\alpha} = \ov{D}^2 D_{\alpha} V'$ and coupling $g'$. It couples to a hidden sector current $J'$ via
\beq
\C{L} \supset \int d^4 \theta g' J' V'.
\eeq
We want to consider the mixing with the visible sector U(1)$_Y$ hypercharge. Therefore, we can define the correlators 
\beq\begin{split}
\bra J (p) J' (-p) \ket & = \tilde{C}_0^{(Y)\prime} (p^2/M^2; M/\Lambda) \\ 
\bra  j_{\alpha} (p) \ov{j}_{\dot{\beta}}^{\prime} (-p) \ket & = -\sigma^{\mu}_{\alpha \dot{\beta}} p_\mu \tilde{C}_{1/2}^{(Y)\prime} (p^2/M^2; M/\Lambda) \\ 
\bra  j_{\mu} (p) \ov{j}_{\nu}^{\prime} (-p) \ket & = -(p^2 \eta_{\mu \nu} - p_\mu p_\nu ) \tilde{C}_{1}^{(Y)\prime} (p^2/M^2; M/\Lambda).
\end{split}\eeq
In the case of supersymmetry, these are equal ($\tilde{C}_0^{(Y)\prime} = \tilde{C}_{1/2}^{(Y)\prime} = \tilde{C}_{1}^{(Y)\prime} = \tilde{C}^{(Y)\prime}$) and we find a term in the Lagrangian
\beq
\C{L} \supset \int d^2 \theta \, \tilde{C}^{(Y)\prime} (p^2) W^{\prime}_{\alpha} W^{\alpha} 
\eeq
but otherwise we generate
\beq
\delta \mathcal{L}_{eff} \supset     g_1 g' \tilde{C}_0^{(Y)\prime} (p^2) \frac{1}{2} D D' - g_1 g' \tilde{C}_{1/2}^{(Y)\prime} (p^2) i \lambda' \sigma^{\mu} \partial_\mu \bar{\lambda} - \frac{1}{4} \tilde{C}_{1}^{(Y)\prime}  (p^2) F_{\mu \nu}' F^{\mu \nu}  +c.c.
\eeq
and therefore for SUSY breaking we are interested in $\tilde{C}_0^{(Y)\prime}$; we find that the formula for squark mass squareds (\ref{MSSsfermion}) should be modified to
\beq
m^2_{\tilde{f}} = g_1^2 Y_f (\xi +   g' \tilde{C}_0^{(Y)\prime} (0) \bra D' \ket )+ \sum_{r=1}^3 g_r^4 C_2 (f;r) A_r.
\eeq
Therefore a D-term for the hidden U(1) manifests itself as a Fayet-Iliopoulos term in the visible one - preserving the mass sum rules. However, since this effect is suppressed by a factor of $g' \tilde{C}_0^{(Y)\prime}$ relative to a tree level FI term it there is no a priori reason to forbid it - and indeed messenger parity does not exclude it. If we assume that $g' \sim g$ and that $\tilde{C}_0^{(Y)\prime} (0) \sim 10^{-9}$ consistent with current limits, then we may have $\bra D' \ket$ as large as $\sim 10^5 GeV^2$, but the size of kinetic mixing may be many orders of magnitude smaller than this, for example if the hidden U(1) is a dark matter constituent \cite{Chen:2008yi,Postma:2008ec,Jaeckel:2008fi,ArkaniHamed:2008qn,Redondo:2008en}. In this case, as pointed out in \cite{Dienes:1996zr}, a kinetic mixing of $10^{-16}$ is consistent with $\bra D' \ket \sim M_Z M_{pl}$. We should thus not discount the possibility that D-term communication via kinetic mixing contributes to the soft masses, yielding a fourth real parameter to the gauge mediation spectrum.

As in the case of the MSS correlators (\ref{MSSCorrelators}), the singular behaviour of the new correlators is given by
\beq
\tilde{C}_a^{(Y)'} (p^2/M^2; M/\Lambda) = 2\pi^2 c \log (\Lambda/M) + \mathrm{finite}
\eeq
with the same $c$ for each, since supersymmetry is restored above the scale $M$. However, for string-derived D-brane models of kinetic mixing where there is no light matter charged under the hidden U(1) \cite{Abel:2003ue,Abel:2006qt,Abel:2008ai}, we generically find that $c=0$; at one loop, it is given by $\tr (Q Q')$, i.e. a trace over the states of the product of their charges under the two U(1)s, and in such models the heavy states satisfy this while still contributing a threshold correction to $\tilde{C}_a^{(Y)'}$.

The general approach to kinetic mixing is useful when we consider that the effect is generated by integrating out super-massive modes, and thus it is sensitive to the UV completion of the theory. If the completion is string theory, then the effect may be generated by either massive bifundamental fields or axions. 
The axionic interaction with gauge fields $V_i$ can be given  by
\beq
\C{L} = \int d^4 \theta \ M_A (\C{A} + \bar{\C{A}}) Q_i V_i 
\eeq
where $\C{A}$ is a chiral multiplet describing the axion with components $(A, s_{\alpha})$ and $Q_i\in \{Q,Q'\,\cdots\}$ are Green-Schwarz coefficients. Writing $a = \Re(A)$ we find $\{J, J'\} = \{M_A Q a, M_A Q' a\}$ and $\{j_{\mu}, j_{\mu}^{\prime}\} = \{M_A Q \partial_{\mu} a,M_A Q^{\prime} \partial_{\mu} a\}$ so
\beq
\tilde{C}_0^{(Y)\prime} (p^2/M^2) = M^2_A Q Q' \bra a (p) a (-p) \ket 
\eeq
and $\tilde{C}_1^{(Y)\prime} = \tilde{C}_0^{(Y)\prime}$ even when supersymmetry is broken. 

\section{Acknowledgments}

Work supported in part by the French ANR contracts
BLAN05-0079-01 and \\
PHYS@COL\&COS. M.~D.~G. is supported by a CNRS contract.

\end{document}